\documentclass[a4paper,11pt]{article}
\usepackage{pos}
\usepackage{lineno}

\title{Dijet acoplanarity as a function of flattenicity in proton-proton collisions at $\sqrt{s}=13$\,TeV}
\ShortTitle{Dijet acoplanarity as a function of flattenicity in pp collisions}

\author*[a]{Jesús Eduardo Muñoz Méndez}
\author[a]{Antonio Ortiz}

\affiliation[a]{Instituto de Ciencias Nucleares, UNAM\\
 Circuito exterior s/n Ciudad Universitaria, 04510 Mexico City, Mexico}

\emailAdd{jesus.munoz@correo.nucleares.unam.mx}

\abstract{One of the most important discoveries at the LHC is the presence of effects associated with quark–gluon plasma (QGP) formation in proton–proton (pp) collisions. The origin of these observations remains unknown. Since event selection in many studies relies on multiplicity-based estimators known to introduce selection biases, alternative approaches must be explored to elucidate the origin of these phenomena.

This contribution presents a comparison of different event classifiers used in LHC experiments, aiming to assess their mutual compatibility. Results for pp collisions at $\sqrt{s}=13$\,TeV simulated with PYTHIA~8 indicate that the different classifiers introduce biases that can obscure any evidence of QGP formation in small systems. However, these biases are significantly reduced or absent in events triggered with flattenicity. This finding motivates the application of flattenicity in dijet acoplanarity studies used to search for jet quenching effects. While multiplicity-based estimators produce an azimuthal broadening in high-multiplicity compared with minimum-bias pp collisions, even in PYTHIA~8 that does not incorporate jet quenching, events triggered with flattenicity show no such broadening. This suggests that this novel tool could be exploited by LHC experiments to refine jet-quenching searches in small systems.}

\FullConference{43rd International Conference on High Energy Physics (ICHEP 2026)\\
30 July  to 5 August , 2026\\
Natal, Brazil\\}

\begin{document}
\maketitle

\section{Introduction}

Jet quenching refers to the energy loss of energetic partons traversing a quark-gluon plasma (QGP), leading to measurable modifications such as suppression of high-$p_{\rm T}$ hadrons and angular deflection of the away-side jet. Although this phenomenon is well established in Pb--Pb collisions through measurements of the nuclear modification factor \cite{ALICE:2018vuu}, there is still no direct evidence in smaller systems such as p--Pb and proton-proton (pp) collisions. Since jet quenching is a direct consequence of QGP formation, its observation in small systems is essential for establishing the minimum system size for QGP-like medium formation. However, searches in pp collisions are challenging for reasons beyond the medium size. Jet quenching studies in pp collisions by ALICE have reported that the observed azimuthal broadening is significantly affected by a selection bias introduced by the event classifier used to define high-activity collisions \cite{ALICE:2023plt}. ALICE concluded that this broadening is mainly caused by this bias rather than by a genuine jet quenching effect. To test whether alternative event classifiers can mitigate this bias, the most widely used classifiers at the LHC were compared using PYTHIA~8 simulations of pp collisions. These include the charged-particle multiplicity at mid-pseudorapidity ($N_{\rm ch}$) and forward pseudorapidity (V0M), as well as transverse sphericity ($S_{\rm T}$)~\cite{ALICE:2012cor, Hanson:1975fe}, spherocity ($S_{0}$)~\cite{Acharya:2019mzb, Cuautle:2015kra}, the relative transverse activity classifier ($R_{\rm T}$)~\cite{ALICE:2023csm,ALICE:2023yuk}, and flattenicity (1-$\rho_{\rm nch}$)~\cite{Ortiz:2022mfv, ALICE:2024vaf}. For each estimator, the top 0.1\% of events, the highest-activity population accessible with that classifier, was selected. The goal was to identify which estimator selects high-activity events without biasing the sample and to understand what kind of collisions are selected. Definitions of each classifier can be found in Ref.~\cite{Mendez:2025dqz}.

\section{$\Delta_{\rm{recoil}}$ observable}

One way to study jet quenching in pp collisions is to examine the azimuthal distribution of charged-particle jets recoiling from a high-$p_{\rm T}$ trigger hadron. In principle, a back-to-back configuration is expected in vacuum, so any deviation from this configuration could indicate a jet quenching effect. This measurement is performed using the $\Delta_{\rm{recoil}}(p_{\rm{T}}, \Delta\varphi)$ observable, defined to eliminate uncorrelated background. This is done by subtracting the distributions corresponding to two different trigger-track (TT) classes, namely the signal distribution $\rm{TT}_{\rm sig}$ and the reference distribution $\rm{TT}_{\rm ref}$:

\begin{equation} 
\Delta_{\rm{recoil}}(p_{\rm{T}}, \Delta\varphi) =
\frac{1}{N_{\rm{trig}}}
\left.
\frac{d^2 N_{\rm{jet}}}
{d p^{\rm{ch}}_{\rm{T,jet}}\, d\Delta\varphi}
\right|_{p_{\rm{T,trig}} \in \rm{TT}_{\rm{Sig}}}
-
c_{\rm{Ref}} \times
\frac{1}{N_{\rm{trig}}}
\left.
\frac{d^2 N_{\rm{jet}}}
{d p^{\rm{ch}}_{\rm{T,jet}}\, d\Delta\varphi}
\right|_{p_{\rm{T,trig}} \in \rm{TT}_{\rm{Ref}}} \, ,
\end{equation}

where $N_\text{trig}$ indicates the corresponding number of trigger hadrons in each TT class. This observable therefore captures the normalized yield of recoil jets as a function of the jet transverse momentum $p^{\rm ch}_{\rm T,jet}$ and the azimuthal difference $\Delta\varphi$ between the trigger hadron and the recoil jet, restricted to the interval $[0, \pi]$. Because hadrons with $p_{\rm T, trig}>20$ GeV/$c$ arise predominantly from hard partonic scatterings at LHC energies, the signal distribution is built from the range $\rm{TT}_{\rm sig}=\rm{TT}\{20,30\}$, i.e. $20<p_{\rm T, trig}<30$ GeV/$c$. In contrast, given that the underlying event saturates at $p_{\rm T, trig}>5$ GeV/$c$ \cite{ALICE:2019mmy}, the reference distribution is constructed from $\rm{TT}_{\rm ref}=\rm{TT}\{6,7\}$, i.e. $6<p_{\rm T, trig}<7$ GeV/$c$. When more than one hadron satisfies these criteria within an event, one is chosen at random to act as the trigger. Since the distribution is semi-inclusive, no requirement is imposed on the recoil jets themselves, so every jet candidate within the accepted range is considered. Meanwhile, the scale factor $c_{\rm Ref}$ compensates for the lower background yield in the $\rm{TT}_{\rm sig}$ distribution compared to $\rm{TT}_{\rm ref}$, since the signal distribution contains a larger fraction of events with correlated recoil jets. This factor is obtained from the ratio of the trigger-normalized signal and reference recoil-jet yields in the range $0<p_{\rm{T,jet}}^{\rm{ch}}<1$~GeV/$c$, which is mostly dominated by the uncorrelated background.

The following results are based on 6 billion inelastic pp collisions at $\sqrt{s}=13$\,TeV (equivalent to an integrated luminosity of around 70\,nb$^{-1}$) simulated with PYTHIA~8 using the Monash tune and HardQCD:all to enable all $2\to2$ QCD scattering processes. The event classifiers listed above are calculated using primary charged particles. For these simulations, minimum bias (MB) events are defined as pp collisions with at least one particle with $p_{\rm T}>0$ GeV/$c$ and within $|\eta|<1$. Jets were reconstructed with the FastJet library \cite{Cacciari:2011ma}, using the $k_{\rm T}$ algorithm to estimate the soft background and the anti-$k_{\rm T}$ algorithm to obtain jet candidates, which were corrected for the calculated background. Both reconstructions used charged particles with $|\eta|<0.9$ and $0.15<p_{\rm T}<100$ GeV/$c$, treated as massless and combined with the boost-invariant $p_{\rm T}$ recombination scheme. The resolution parameter was set to $R=0.4$ so that only jets within the central barrel were accepted. The two hardest jets in the event were excluded, and the ghost area was 0.005.

\begin{figure}[t]
\centering
\includegraphics[width=0.5\textwidth]{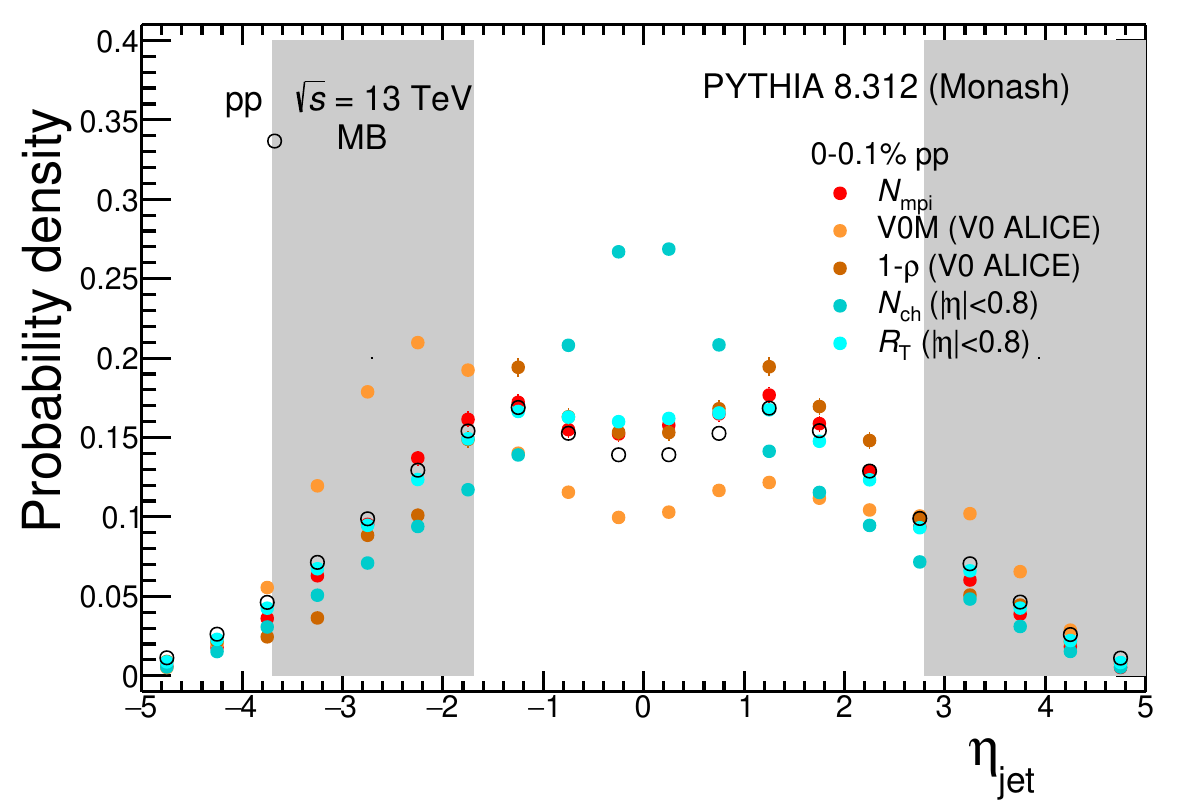}
\caption{Probability distribution of jets recoiling from a high-$p_{\rm T}$ hadron (TT\{20,30\} and $|\eta|<0.8$) as a function of $\eta_{\rm jet}$ for $p_{\rm T}^{\rm jet}>25$ GeV/$c$, in pp collisions at $\sqrt{s}=13$\,TeV. Results for various event-activity estimators are shown in the 0-0.1\% event class. The shaded areas represent the pseudorapidity regions covered by the ALICE V0 detector. The error bars represent the statistical uncertainty. Figure taken from Ref.~\cite{Mendez:2025dqz}.}
\label{fig:6}  
\end{figure}

\section{Bias toward local fluctuations}

Figure \ref{fig:6} shows the distribution of jets recoiling from a high-$p_{\rm T}$ leading hadron as a function of the jet pseudorapidity ($\eta_{\rm jet}$) for the 0-0.1\% event-activity class selected using the different estimators. The shaded areas indicate the regions covered by the ALICE V0 detector. The distributions for MB events and for a selection based on the exact number of multi-parton interactions ($N_{\rm mpi}$), as provided by the MC generator, are also included. The distributions of events triggered with sphericity and spherocity are absent from the figure since most of their recoil jets did not satisfy the selection criteria. As shown previously by ALICE \cite{ALICE:2023plt}, the V0M distribution exhibits an asymmetry, indicating a preferential selection of events with a recoil jet in the V0C acceptance. Meanwhile, the remaining event classifiers yield symmetric distributions. However, $N_{\rm ch}$ shows an enhancement in the pseudorapidity region over which it is calculated ($|\eta|<0.8$), indicating an autocorrelation or a bias towards coplanar jets. Notably, the distributions for events selected using flattenicity and $R_{\rm T}$  closely resemble those for MB events and the MC classifier ($N_{\rm mpi}$).

\section{Bias toward harder-than-average events}

\begin{figure}[t]
\centering
\begin{minipage}[t]{0.48\textwidth}
\centering
\includegraphics[width=\linewidth]{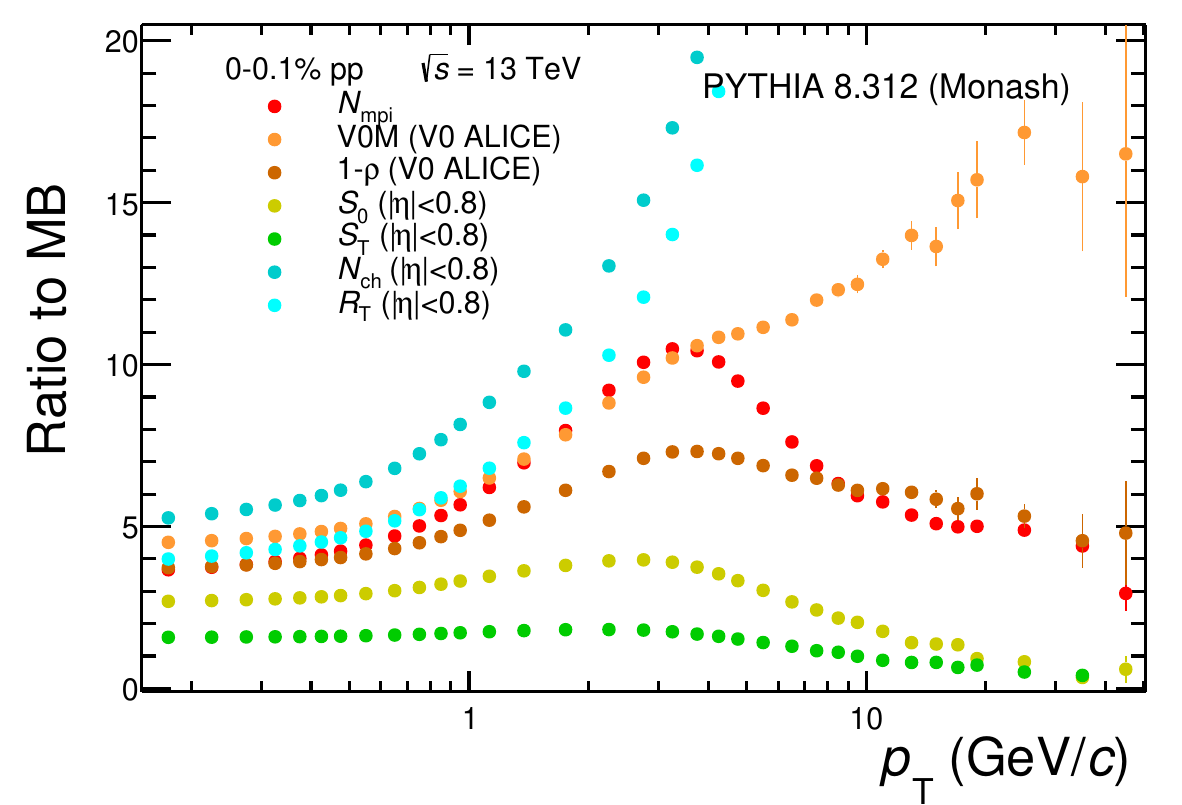}
\captionof{figure}{Transverse momentum spectra for the 0-0.1\% event class normalized to the minimum-bias $p_{\rm T}$ spectrum. The error bars represent the statistical uncertainty. Figure taken from Ref.~\cite{Mendez:2025dqz}.}
\label{fig:5}
\end{minipage}\hfill
\begin{minipage}[t]{0.48\textwidth}
\centering
\includegraphics[width=\linewidth]{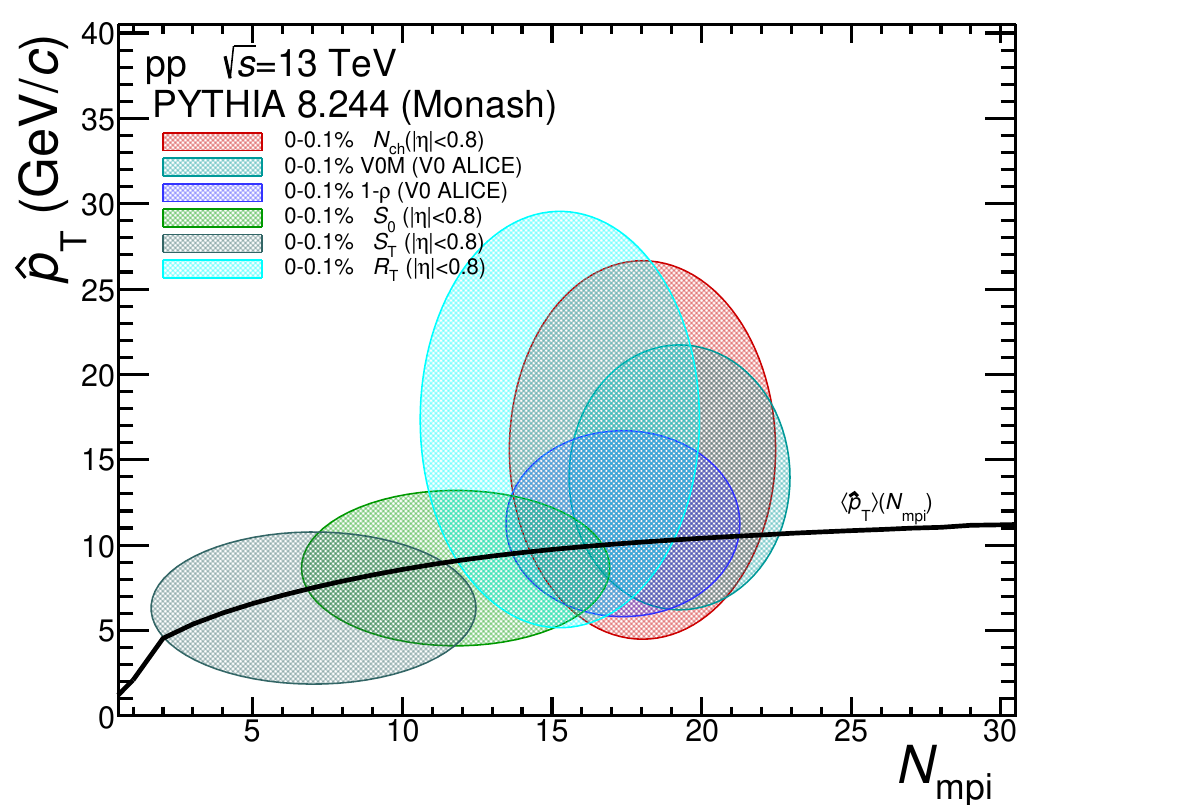}
\captionof{figure}{Phase space in the $\hat{p}_{\rm T}$-$N_{\rm mpi}$ plane covered by the different 0-0.1\% event classes. Figure taken from Ref.~\cite{Mendez:2025dqz}.}
\label{fig:8}
\end{minipage}
\end{figure}

Another way to illustrate the contrasting event selections produced by the classifiers is through the transverse-momentum spectrum for the 0-0.1\% event class, normalized to the spectrum in minimum-bias pp collisions, as shown in Fig.~\ref{fig:5}. While a clear suppression of high-$p_{\rm T}$ particles can be observed for sphericity and spherocity, multiplicity-dependent classifiers at mid-pseudorapidity ($N_{\rm ch}$ and $R_{\rm T}$) instead exhibit a bias towards harder-than-average pp collisions. This bias is slightly reduced in the case of the V0M estimator. However, at higher $p_{\rm T}$, the result obtained with flattenicity is more consistent with that obtained using the $N_{\rm mpi}$ selection. 

A final comparison of these classifiers is shown in Fig.~\ref{fig:8}, which displays the one-sigma region in the $\hat{p}_{\rm T}$-$N_{\rm mpi}$ plane covered by each of the 0-0.1\% event classes, where $\hat{p}_{\rm T}$ indicates the transverse momentum of the main parton-parton scattering. The colored bands show the range selected by the different event classifiers, while the solid line shows the expected trend: the average transverse momentum of the main scattering increases with $N_{\rm mpi}$, since more parton-parton interactions raise the probability that at least one is a hard scattering. Two clear families can be observed in the figure. The first corresponds to multiplicity-based classifiers ($R_{\rm T}$, $N_{\rm ch}$ and V0M), which have a very broad $\hat{p}_{\rm T}$ range, indicating their tendency to select harder-than-average pp collisions. In contrast, event-shape classifiers $(S_{\rm T}, S_{\rm 0}$ and $1-\rho_{\rm nch})$ exhibit a behavior closer to the average. Thus, even though every classifier selects the most active events, the overlap among the selected event populations is rather small. Therefore, results obtained with one classifier cannot be directly compared to results from another. They are effectively probing different slices of the collision landscape.

\section{Dijet acoplanarity as a function of flattenicity}

\begin{figure}[t]
\centering
\includegraphics[width=0.45\textwidth]{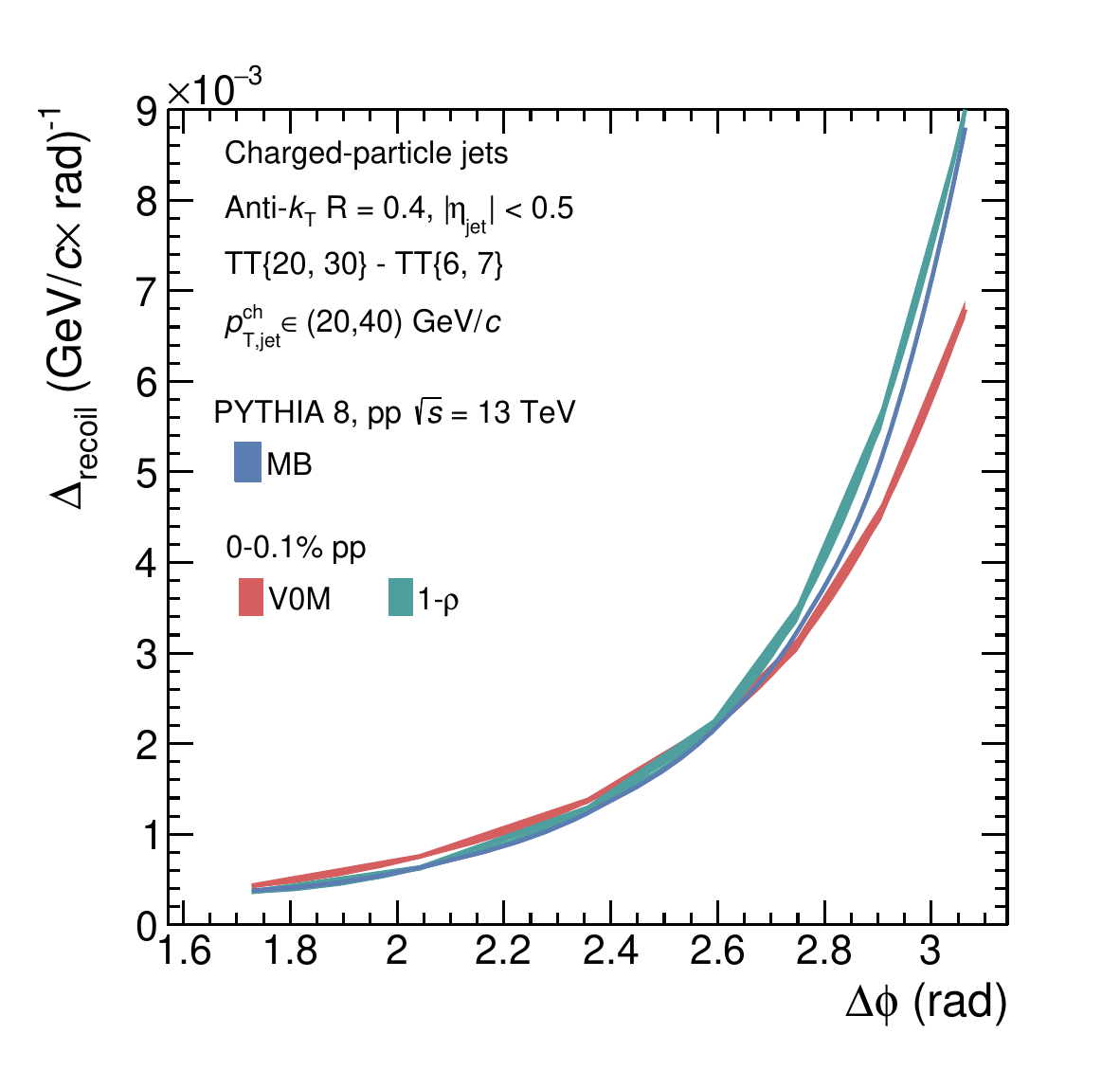}
\caption{PYTHIA~8 results for the $\Delta_{\rm recoil}(\Delta\varphi)$ distributions in the range $20<p^{\rm{ch}}_{\rm{T,jet}}<40$ $\rm{GeV}/c$ for MB and the 0-0.1\% event class selected with V0M and flattenicity, in pp collisions at $\sqrt{s}=13$ TeV. The width of the bands represents the statistical uncertainty.}
\label{fig:delta_phi}
\end{figure}

The previous results showed that, unlike V0M, flattenicity reduces the bias reported by ALICE. This motivates the use of flattenicity in $\Delta_{\rm{recoil}}(\Delta\varphi)$ measurements to test whether the reported broadening observed in dijet studies persists once this bias is removed. Figure~\ref{fig:delta_phi} shows the resulting $\Delta_{\rm{recoil}}(\Delta\varphi)$ distribution obtained with PYTHIA~8 for MB and high-activity events selected using either flattenicity ($1-\rho_{\rm nch}$) or V0M, in pp collisions at $\sqrt{s}=13$ TeV. The result indicates that the distribution for events selected using flattenicity is fully consistent with the MB distribution and, therefore, no broadening is observed when this classifier is considered. Thus, within PYTHIA 8, the broadening observed with the V0M selection can be attributed to the reported selection bias. Since flattenicity minimizes jet fragmentation biases, any future broadening observed with this classifier would provide a stronger indication of genuine medium-induced jet quenching.

\section{Conclusions}

A comparison among event classifiers commonly used at the LHC was performed using PYTHIA~8 simulations of pp collisions at $\sqrt{s}=13$\,TeV. The results indicate that multiplicity-based estimators ($N_{\rm ch}$, V0M and $R_{\rm T}$) favor harder-than-average collisions and local fluctuations, while sphericity and spherocity suppress high-$p_{\rm T}$ particles and recoil jets. Flattenicity significantly reduces these biases and, when applied to the $\Delta_{\rm recoil}(\Delta\varphi)$ observable, shows no azimuthal broadening, in contrast to the broadening obtained with V0M. This makes flattenicity a promising tool to refine jet quenching searches in small systems.\\

\noindent\textbf{Acknowledgments.} This work has been supported by DGAPA-UNAM PAPIIT No. IG100524

\end{document}